\documentclass[sigconf]{acmart}  

\usepackage{bbm}  

\usepackage{amsmath}          
\usepackage{amssymb} 			    
\usepackage{textcomp}

\usepackage{url}              

\usepackage{array}            
\usepackage{tabularx}         
\usepackage{colortbl}
\newcolumntype{Y}{>{\centering\arraybackslash}X}	

\usepackage{mathtools}          

\DeclareMathOperator*{\argmin}{arg\,min} 

\usepackage{pbox}   

\usepackage{setspace}

\usepackage{algorithm}
\usepackage[noend]{algpseudocode} 

\algnewcommand{\COMMENT}[2][.5\linewidth]{\leavevmode\hfill\makebox[#1][l]{//~#2}}
\algnewcommand{\LineComment}[1]{\State \(//\) #1}	
\algnewcommand\RETURN{\State \textbf{return} }

\usepackage{graphicx}
\usepackage[space]{grffile}   
\graphicspath{{.}}
\DeclareGraphicsExtensions{.pdf,.jpeg,.jpg,.png}

\usepackage[font={bf}]{caption}
\usepackage{paralist}
\setdefaultleftmargin{10pt}{10pt}{}{}{}{}

\usepackage{outlines}
\usepackage{enumitem}
\newcommand{\ItemSpacing}{0mm}
\newcommand{\ParSpacing}{0mm}
\setenumerate[1]{itemsep={\ItemSpacing},parsep={\ParSpacing},label=\arabic*.}
\setenumerate[2]{itemsep={\ItemSpacing},parsep={\ParSpacing}}

\usepackage[linewidth=1pt]{mdframed}
\mdfsetup{frametitlealignment=\center, skipabove=0, innertopmargin=1mm,
innerleftmargin=2mm, leftmargin=0mm, rightmargin=0mm}

\newlength\myindent
\renewcommand{\vec}[1]{\mathbf{#1}}
\newcommand{\mat}[1]{\mathbf{#1}}

\DeclarePairedDelimiter\ceil{\lceil}{\rceil}
\DeclarePairedDelimiter\floor{\lfloor}{\rfloor}

\newcommand\eps\varepsilon

\renewcommand\inf\infty

\newcommand{\q}{\vec{q}}
\renewcommand{\r}{\vec{r}}
\newcommand{\x}{\vec{x}}
\newcommand{\xhat}{\hat{\vec{x}}}

\DeclarePairedDelimiter\abs{\lvert}{\rvert}%
\DeclarePairedDelimiter\norm{\lVert}{\rVert}%

\DeclareMathOperator{\Var}{Var}

\begin{document}

\setcopyright{rightsretained}

\copyrightyear{2017}
\acmYear{2017}
\setcopyright{acmlicensed}
\acmConference{KDD'17}{}{August 13--17, 2017, Halifax, NS, Canada.} \acmPrice{15.00} \acmDOI{10.1145/3097983.3098195}
\acmISBN{ISBN 978-1-4503-4887-4/17/08}

\fancyhead{}  

\title{Bolt: Accelerated Data Mining with Fast Vector Compression}

\author{Davis W. Blalock}
\affiliation{
  \institution{Computer Science and Artificial \\ Intelligence Laboratory}
  \institution{Massachusetts Institute of Technology}
}
\email{dblalock@mit.edu}

\author{John V. Guttag}
\affiliation{
  \institution{Computer Science and Artificial \\ Intelligence Laboratory}
  \institution{Massachusetts Institute of Technology}
}
\email{guttag@mit.edu}

\begin{abstract}

Vectors of data are at the heart of machine learning and data mining. Recently, vector quantization methods have shown great promise in reducing both the time and space costs of operating on vectors. We introduce a vector quantization algorithm that can compress vectors over $12\times$ faster than existing techniques while also accelerating approximate vector operations such as distance and dot product computations by up to $10\times$. Because it can encode over 2GB of vectors per second, it makes vector quantization cheap enough to employ in many more circumstances. For example, using our technique to compute approximate dot products in a nested loop can multiply matrices faster than a state-of-the-art BLAS implementation, even when our algorithm must first compress the matrices.

In addition to showing the above speedups, we demonstrate that our approach can accelerate nearest neighbor search and maximum inner product search by over $100\times$ compared to floating point operations and up to $10\times$ compared to other vector quantization methods. Our approximate Euclidean distance and dot product computations are not only faster than those of related algorithms with slower encodings, but also faster than Hamming distance computations, which have direct hardware support on the tested platforms. We also assess the errors of our algorithm's approximate distances and dot products, and find that it is competitive with existing, slower vector quantization algorithms. 

\end{abstract}



\maketitle

\section{Introduction} \label{sec:intro}



As datasets grow larger, so too do the costs of mining them. These costs include not only the space to store the dataset, but also the compute time to operate on it. This time cost can be decomposed into:
\begin{align}
    \texttt{Cost}_{\text{time}} = \texttt{Cost}_{\text{read}} + \texttt{Cost}_{\text{write}}
\end{align}
where $\texttt{Cost}_{\texttt{read}}$ is the time cost of operations that read the data, and $\texttt{Cost}_{\texttt{write}}$ is the time cost of creating, updating, or deleting data.

For datasets of vectors, for which many of the read operations are scalar reductions such as Euclidean distance and dot product computations, vector quantization methods enable significant savings in both space usage and $\texttt{Cost}_{\texttt{read}}$. By replacing each vector with a learned approximation, these methods both save space and enable fast approximate distance and similarity computations. With as little as 8B per vector, these techniques can often preserve distances and dot products with extremely high accuracy \cite{lsq,otq,opq,cq,stackedQuantizers}.




However, computing the approximation for a given vector can be time-consuming, adding greatly to $\texttt{Cost}_{\texttt{write}}$. The state-of-the-art method of \cite{lsq}, for example, requires up to \textit{4ms} to encode a single $128$-dimensional vector. This makes it practical only if there are few writes per second. Other techniques are faster, but as we show experimentally, there is significant room for improvement.

We describe a vector quantization algorithm, \textit{Bolt}, that greatly reduces both the time to encode vectors ($\texttt{Cost}_{\texttt{write}}$) and the time to compute scalar reductions over them ($\texttt{Cost}_{\texttt{read}}$). This not only reduces the overhead of quantization, but also increases its benefits, making it worthwhile for many more datasets. Our key ideas are to 1) learn an approximation for the lookup tables used to compute scalar reductions; and 2) use much smaller quantization codebooks than similar techniques. Together, these changes facilitate finding optimal vector encodings and allow scans over codes to be done in a computationally vectorized manner.

Our contributions consist of:
\begin{enumerate}
\item A vector quantization algorithm that encodes vectors significantly faster than existing algorithms for a given level of compression.
\item A fast means of computing approximate similarities and distances using quantized vectors. Possible similarities and distances include dot products, cosine similarities, and distances in $L_p$ spaces, such as the Euclidean distance.
\end{enumerate}










\subsection{Problem Statement}


Let $\vec{q} \in \mathbb{R}^J$ be a \textit{query} vector and let $\mathcal{X} = \{\vec{x}_1,\ldots,\vec{x}_N\}, \vec{x}_i \in \mathbb{R}^J$ be a collection of \textit{database} vectors. Further let $d: \mathbb{R}^J \times \mathbb{R}^J \rightarrow \mathbb{R}$ be a distance or similarity function that can be written as:
\begin{align} \label{eq:distFuncForm}
        d(\vec{q}, \vec{x}) = f \big( \sum_{j=1}^J \delta(q_j, x_j) \big)
\end{align}


where $f: \mathbb{R} \rightarrow \mathbb{R}$, $\delta: \mathbb{R} \times \mathbb{R} \rightarrow \mathbb{R}$. This includes both distances in $L_p$ spaces and dot products as special cases. In the former case, $\delta(q_j, x_j) = |q_j - x_j|^p$ and $f(r) = r^{(1/p)}$; in the latter case, $\delta(q_j, x_j) = q_j x_j$ and $f(r) = r$. For brevity, we will henceforth refer to $d$ as a distance function and its output as a distance, though our remarks apply to all functions of the above form unless otherwise noted.

Our task is to construct three functions $g: \mathbb{R}^J \rightarrow \mathcal{G}$, $h: \mathbb{R}^J \rightarrow \mathcal{H}$, and $\hat{d}: \mathcal{G} \times \mathcal{H} \rightarrow \mathbb{R}$ such that for a given approximation loss $\mathcal{L}$,
\begin{align}
    \mathcal{L} = E_{\vec{q},\vec{x}}[(d(\vec{q}, \vec{x}) - \hat{d}(g(\vec{q}), h(\vec{x})))^2]
\end{align}
the computation time $T$,
\begin{align}
    T = T_g + T_h + T_d
\end{align}
is minimized, where $T_g$ is the time to encode received queries $\vec{q}$ using $g$,\footnote{We cast the creation of query-specific lookup tables as encoding $\vec{q}$ rather than creating a new $\hat{d}$ (the typical interpretation in recent literature).} $T_h$ is the time to encode the database $\mathcal{X}$ using $h$, and $T_d$ is the time to compute the approximate distances between the encoded queries and encoded database vectors. The relative contributions of each of these terms depends on how frequently $\mathcal{X}$ changes, so many of our experiments characterize each separately. 



There is a tradeoff between the value of the loss $\mathcal{L}$ and the time $T$, so multiple operating points are possible. In the extreme cases, $\mathcal{L}$ can be fixed at 0 by setting $g$ and $h$ to identity functions and setting $\hat{d} = d$. Similarly, $T$ can be set to $0$ by ignoring the data and estimating $d$ as a constant. The primary contribution of this work is therefore the introduction of $g$, $h$ and $\hat{d}$ functions that are significantly faster to compute than those of existing work for a wide range of operating points.

\subsection{Assumptions}

Like other vector quantization work \cite{pq,lsq,otq,opq,cq}, we assume that there is an initial offline phase during which the functions $g$ and $h$ may be learned. This phase contains a training dataset for $\mathcal{X}$ and $\vec{q}$. Following this offline phase, there is an online phase wherein we are given database vectors $\vec{x}$ that must be encoded and query vectors $\vec{q}$ for which we must compute the distances to all of the database vectors received so far. Once a query is received, these distances must be computed with as little latency as possible. The vectors of $\mat{X}$ may be given all at once, or one at a time; they may also be modified or deleted, necessitating re-encoding or removal. This is in contrast to most existing work, which assumes that $\vec{x}$ vectors are all added at once before any queries are received \cite{pq,lsq,otq,opq,cq}, and therefore that encoding speed is less of a concern.

In practice, one might require the distances between $\vec{q}$ and only some of the database vectors $\mathcal{X}$ (in particular, the $k$ closest vectors). This can be achieved using an indexing structure, such as an Inverted Multi-Index \cite{IMI, NOIMI} or Locality-Sensitive Hashing hash tables \cite{E2LSH, crosspolytopeLSH}, that allow inspection of only a fraction of $\mathcal{X}$. Such indexing is complementary to our work in that our approach could be used to accelerate the computation of distances to the subset of $\mathcal{X}$ that is inspected. Consequently, we assume that the task is to compute the distances to all vectors, noting that, in a production setting, ``all vectors'' for a given query might be a subset of a full database.

Finally, we assume that both $\mathcal{X}$ and $\vec{q}$ are relatively dense. Bolt can be applied to sparse data, but does not leverage the sparsity. Consequently, it is advisable to embed sparse vectors into a dense, lower-dimensional space before using Bolt.


\vspace{-2mm}




\section{Related Work} \label{sec:relatedWork}

Accelerating vector operations through compression has been the subject of a great deal of research in the computer vision, information retrieval, and machine learning communities, among others. Our review will necessarily be incomplete, so we refer the reader to \cite{learningToHashSurvey, hashingSimilaritySurvey} for detailed surveys.

Many existing approaches in the computer vision and information retrieval literature fall into one of two categories \cite{hashingSimilaritySurvey}: binary embedding and vector quantization. Binary embedding techniques seek to map vectors in $\mathbb{R}^J$ to $B$-dimensional Hamming space, typically with $B < J$. The appeal of binary embedding is that a $B$-element vector in Hamming space can be stored in $B$ bits, affording excellent compression. Moreover, the \texttt{popcount} instruction present on virtually all desktop, smart phone, and server processors can be used to compute Hamming distances between 8 byte vectors in as little as three cycles. This fast distance computation comes at the price of reduced representational accuracy for a given code length \cite{opq,hashingSimilaritySurvey}. He et al. \cite{opq} showed that the popular binary embedding technique of \cite{ITQ} is a more constrained version of their vector quantization algorithm, and that the objective function of another state-of-the art binary embedding \cite{isohash} can be understood as maximizing only one of two sufficient conditions for optimal encoding of Gaussian data.

Vector quantization approaches yield lower errors than binary embedding for a given code length, but entail slower encoding and distance computations. The simplest and most popular vector quantization method is $k$-means, which can be seen as encoding a vector as the centroid to which it is closest. A generalization of $k$-means, Product Quantization (PQ) \cite{pq}, splits the vector into $M$ disjoint subvectors and runs $k$-means on each. The resulting code is the concatenation of the codes for each subspace. Numerous generalizations of PQ have been published, including Cartesian $k$-means \cite{cartesianKmeans}, Optimized Product Quantization \cite{opq},  Generalized Residual Vector Quantization \cite{grvq}, Additive Quantization \cite{aq}, Composite Quantization \cite{cq}, Optimized Tree Quantization \cite{otq}, Stacked Quantizers \cite{stackedQuantizers}, and Local Search Quantization \cite{lsq}. The idea behind most of these generalizations is to either rotate the vectors or relax the constraint that the subvectors be disjoint. Collectively, these techniques that rely on using the concatenation of multiple codes to describe a vector are known as Multi-Codebook Quantization (MCQ) methods. 

An interesting hybrid of binary embedding and vector quantization is the recent Polysemous Coding of Douze et al. \cite{polysemous}. This encoding uses product quantization codebooks optimized to also function as binary codes, allowing the use of Hamming distances as a fast approximation that can be refined for promising nearest neighbor candidates. 

The most similar vector quantization-related algorithm to our own is that of \cite{simdpq}, which also vectorizes PQ distance computations. However, their method requires hundreds of thousands or millions of encodings to be sorted lexicographically and stored contiguously ahead of time, as well as scanned through serially. This is not possible when the data is rapidly changing or when using an indexing structure, which would split the data into far smaller partitions. Their approach also requires a second refinement pass of non-vectorized PQ distance computations, making their reported speedups significantly lower than our own.

In the machine learning community, accelerating vector operations has been done primarily through (non-binary) embedding, structured matrices, and model compression. Embedding yields acceleration by reducing the dimensionality of data while preserving the relevant structure of a dataset overall. There are strong theoretical guarantees regarding the level of reduction attainable for a given level of distortion in pairwise distances \cite{jl, fastJL, jlIsTight}, as well as strong empirical results \cite{superBitLSH, compressiveMining}. However, because embedding \textit{per se} only entails reducing the number of floating-point numbers stored, without reducing the size of each, it is not usually competitive with vector quantization methods. It is possible to embed data before applying vector quantization, so the two techniques are complementary.

An alternative to embedding that reduces the cost of storing and multiplying by matrices is the use of structured matrices. This consists of repeatedly applying a linear transform, such as permutation \cite{adaptiveFastfood}, the Fast Fourier Transform \cite{orthogonalRandomFeatures}, the Discrete Cosine Transform \cite{acdc}, or the Fast Hadamard Transform \cite{crosspolytopeLSH, structuredSpinners}, possibly with learned elementwise weights, instead of performing a matrix multiply. These methods have strong theoretical grounding \cite{structuredSpinners} and sometimes outperform non-structured matrices \cite{adaptiveFastfood}. They are orthogonal to our work in that they bypass the need for a dense matrix entirely, while our approach can accelerate operations for which a dense matrix is used.

Another vector-quantization-like technique common in machine learning is model compression. This typically consists of some combination of 1) restricting the representation of variables, such as neural network weights, to fewer bits \cite{lossAwareDnnBinarization}; 2) reusing weights \cite{hashnet}; 3) pruning weights in a model after training \cite{diversityNets, learningWeightsConnections}; and 4) training a small model to approximate the outputs of a larger model \cite{rnnDarkKnowledge}. This has been a subject of intense research for neural networks in recent years, so we do not believe that our approach could yield smaller neural networks than the current state of the art. Instead, our focus is on accelerating operations on weights and data that would otherwise not have been compressed.

\section{Method} \label{sec:method}

As mentioned in the problem statement, our goal is to construct a distance function $\hat{d}$ and two encoding functions $g$ and $h$ such that $\hat{d}(g(\vec{q}), h(\vec{x})) \approx d(\vec{q}, \vec{x})$ for some ``true'' distance function $d$. To explain how we do this, we first begin with a review of Product Quantization \cite{pq}, and then describe how our method differs.

\subsection{Background: Product Quantization}

Perhaps the simplest form of vector quantization is the $k$-means algorithm, which quantizes a vector to its closest centroid among a fixed \textit{codebook} of possibilities. As an encoding function, it transforms a vector into a $\ceil{log_2(K)}$-bit \textit{code} indicating which centroid is closest, where $K$ is the codebook size (i.e., number of centroids). Using this encoding, the distance between a query and a database vector can be approximated as the distance between the query and its associated centroid.

Product Quantization (PQ) is a generalization of $k$-means wherein the vector is split into disjoint \textit{subvectors} and the full vector is encoded as the concatenation of the codes for the subvectors. Then, the full distance is approximated as the sum of the distances between the subvectors of $\vec{q}$ and the chosen centroids for each corresponding subvector of $\vec{x}$.

Formally, PQ approximates the function $d$ as follows. First, recall that, by assumption, $d$ can be written as:
\begin{align*}
        d(\vec{q}, \vec{x}) = f \big( \sum_{j=1}^J \delta(q_j, x_j) \big)
\end{align*}
where $f: \mathbb{R} \rightarrow \mathbb{R}$, \hspace{.5mm} $\delta: \mathbb{R} \times \mathbb{R} \rightarrow \mathbb{R}$. Now, suppose one partitions the indices $j$ into $M$ disjoint subsets $\{p_1,\ldots,p_M \}$. Typically, each subset is a sequence of $J/M$ consecutive indices. 
The argument to $f$ can then be written as:
\begin{align}
        \sum_{m=1}^M \sum_{j \in p_m} \delta(q_j, x_j)
            = \sum_{m=1}^M \boldsymbol{\delta} \left( \vec{q}^{(m)}, \vec{x}^{(m)} \right)
\end{align}
where $\vec{q}^{(m)}$ and $\vec{x}^{(m)}$ are the \textit{subvectors} formed by gathering the elements of $\vec{q}$ and $\vec{x}$ at the indices $j \in p_m$, and $\boldsymbol{\delta}$ sums the $\delta$ functions applied to each dimension. Product quantization replaces each $\vec{x}^{(m)}$ with one vector $\vec{c}_i^{(m)}$ from a \textit{codebook} set $\mathcal{C}_m$ of possibilities. That is: 
\begin{align} \label{eq:pqDistNoLut}
        \sum_{m=1}^M \boldsymbol{\delta} \left( \vec{q}^{(m)}, \vec{x}^{(m)} \right) \approx \sum_{m=1}^M \boldsymbol{\delta} \left( \vec{q}^{(m)}, \vec{c}_i^{(m)} \right)
\end{align}
This allows one to store only the index of the codebook vector chosen (i.e., $i$), instead of the elements of the original vector $\vec{x}^{(m)}$. More formally, let $\mathcal{\mat{C}} = \{\mathcal{C}_1,\ldots,\mathcal{C}_M\}$ be a set of $M$ codebooks where each codebook $\mathcal{C}_m $ is itself a set of $K$ vectors $\{\vec{c}_{1}^{(m)},\ldots,\vec{c}_{K}^{(m)}\}$; we will refer to these vectors as \textit{centroids}. Given this set of codebooks, the PQ encoding function $h(\vec{x})$ is:
\begin{align}
    h(\vec{x}) = [ i_1 \hspace{.25mm}; \hspace{.25mm} \ldots \hspace{.25mm}; \hspace{.25mm} i_M ], \hspace{1mm} i_m = \argmin_i d \left( \vec{c}_i^{(m)}, \vec{x}^{(m)} \right)
\end{align}
That is, $h(\vec{x})$ is a vector such that $h(\vec{x})_m$ is the index of the centroid within codebook $m$ to which $\vec{x}^{(m)}$ is closest.

Using these codebooks also enables construction of a fast query encoding $g$ and distance approximation $\hat{d}$. Specifically, let the query encoding space $\mathcal{G}$ be $R^{K \times M}$ and define $\mat{D} = g(\vec{q})$ as: 
\begin{align}
    \mat{D}_{im} \triangleq \boldsymbol{\delta} \left( \vec{q}^{(m)}, \vec{c}_i^{(m)} \right)
\end{align}
Then we can rewrite the approximate distance on the right hand side of ~\ref{eq:pqDistNoLut} as:
\begin{align} \label{eq:pqDist}
        \sum_{m=1}^M \mat{D}_{im}, \hspace{2mm} i = h(\vec{x})_m
\end{align}
In other words, the distance can be reduced to a sum of precomputed distances between $\vec{q}^{(m)}$ and the codebook vectors $\vec{c}_i^{(m)}$ used to approximate $\vec{x}$. Each of the $M$ columns of $\mat{D}$ represents the distances between $\vec{q}^{(m)}$ and the $K$ centroids in codebook $\mathcal{C}_M$. Computation of the distance proceeds by iterating through the columns, looking up the distance in row $h(\vec{x})_m$, and adding it to a running total. By reintroducing $f$, one can now define:
\begin{align} \label{eq:pq_dhat}
    \hat{d}(g(\vec{q}), h(\vec{x})) \triangleq f \big( \sum_{m=1}^M \mat{D}_{im}, i = h(\vec{x})_m \big)
\end{align}
If $M \ll D$ and $K \ll |\mathcal{X}|$, then computation of $\hat{d}$ is much faster than computation of $d$ given the $g(\vec{q})$ matrix $\mat{D}$ and data encodings $\mathcal{H} = \{h(\vec{x}), \vec{x} \in \mathcal{X} \}$.

The total computational cost of product quantization is $\Theta(KJ)$ to encode each $\vec{x}$, $\Theta(KJ)$ to encode each query $\vec{q}$, and $\Theta(M)$ to compute the approximate distance between an encoded $\vec{q}$ and encoded $\vec{x}$. Because queries must be encoded before distance computations can be performed, this means that the cost of computing the distances to the $N$ database vectors $\mathcal{X}$ when a query is received is $\Theta(KJ) + \Theta(NM)$. Lastly, since codebooks are learned using $k$-means clustering, the time to learn the codebook vectors is $O(KNJT)$, where $T$ is the number of $k$-means iterations. In all works of which we are aware, $K$ is set to $256$ so that each element of $h(\vec{x})$ can be encoded as one byte. 

In certain cases, product quantization is nearly an optimal encoding scheme. Specifically, under the assumptions that:
\begin{enumerate}[leftmargin=7mm]
\item $\vec{x} \sim MVN(\vec{\mu}, \mat{\Sigma})$, and therefore $\vec{x}^{(m)} \sim MVN(\vec{\mu}_m, \mat{\Sigma}_m)$,
\item $\forall_m |\mat{\Sigma}_m| = |\mat{\Sigma}|^{1/m}$,
\end{enumerate}
PQ achieves the information-theoretic lower bound on code length for a given quantization error \cite{opq}. This means that PQ encoding is optimal if $\vec{x}$ is drawn from a multivariate Gaussian and the subspaces $p_m$ are independent and have covariance matrices with equal determinants.

In practice, however, most datasets are not Gaussian and their subspaces are neither independent nor described by similar covariances. Consequently, many works have generalized PQ to capture relationships across subspaces or decrease the dependencies between them \cite{opq, cartesianKmeans, aq, otq, lsq}.

In summary, PQ consists of three components:
\begin{enumerate}[leftmargin=7mm]
    \item Encoding each $\vec{x}$ in the database using $h(\vec{x})$. This transforms $\vec{x}$ to a list of $M$ 8-bit integers, representing the indices of the closest centroids in the $M$ codebooks.
    \item Encoding a query $\vec{q}$ when it is received using $g(\vec{q})$. This returns a $K \times M$ matrix $\mat{D}$ where the $m$th column is the distances to each centroid in codebook $\mathcal{C}_m$.
    \item Scanning the database. Once a query is computed, the approximate distance to each $\vec{x}$ is computed using~(\ref{eq:pq_dhat}) by looking up and summing the appropriate entries from each column of $\mat{D}$.
\end{enumerate}

\subsection{Bolt}

Bolt is similar to product quantization but differs in two key ways:
\begin{enumerate}[leftmargin=7mm]
\item It uses much smaller codebooks.
\item It approximates the distance matrix $\mat{D}$.
\end{enumerate}

Change (1) directly increases the speeds of the encoding functions $g$ and $h$. This is because it reduces the number of $k$-means centroids for which the distances to a given subvector $\vec{x}^{(m)}$ or $\vec{q}^{(m)}$ must be computed. More specifically, by using $K = 16$ centroids (motivated below) instead of 256, we reduce the computation by a factor of $256 / 16 = 16$. This is the source of Bolt's fast encoding. Using fewer centroids also reduces the $k$-means training time, although this is not our focus.

Change (2), approximating the query distance matrix $\mat{D}$, allows us to reduce the size of $\mat{D}$. This approximation is separate from approximating the overall distance---in other algorithms, the entries of $\mat{D}$ are the exact distances between each $\vec{q}^{(m)}$ and the corresponding centroids $\mathcal{C}_m$. In Bolt, the entries of $\mat{D}$ are learned 8-bit quantizations of these exact distances.

Together, changes (1) and (2) allow hardware vectorization of the lookups in $\mat{D}$. Instead of looking up the entry in a given column of $D$ for one $\vec{x}$ (a standard load from memory), we can leverage vector instructions to instead perform $V$ lookups for $V$ consecutive $h(\vec{x})$, $h(\vec{x}_i),\ldots,h(\vec{x}_{i+V})$, where $V = $ 16, 32, or 64 depending on the platform. Under the mild assumption that encodings can be stored in blocks of at least $V$ elements, this affords roughly a $V$-fold speedup in the computation of distances. The ability to perform such vectorized lookups is present on nearly all modern desktops, laptops, servers, tablets, and CUDA-enabled GPUs.\footnote{The relevant instructions are \texttt{vpshufb} on x86, \texttt{vtbl} on ARM, \texttt{vperm} on PowerPC, and \texttt{\_\_shfl} on CUDA.} Consequently, while the performance gain comes from fairly low-level hardware functionality, Bolt is not tied to any particular architecture, processor, or platform.

Mathematically, the challenge in the above approach is quantizing $\mat{D}$. The distances in this matrix vary tremendously as a function of dataset, query vector, and even codebook. Naively truncating the floating-point values to integers in the range [0, 255], for example, would yield almost entirely 0s for datasets with entries $ \ll 1$ and almost entirely 255s for datasets with entries $ \gg 255$. This can of course be counteracted to some extent by globally shifting and scaling the dataset, but such global changes do not account for query-specific and codebook-specific variation.

Consequently, we propose to learn a quantization function at training time. The basic approach is to learn the distribution of distances within a given column of $\mat{D}$ (the distances to centroids within one codebook) across many queries sampled from the training set and find upper and lower cutoffs such that the expected squared error between the quantized and original distances is minimized.

Formally, for a given column $m$ of $\mat{D}$ (henceforth, one \textit{lookup table}), let $Q$ be the distribution of query subvectors $\vec{q}^{(m)}$, $X$ be the distribution of database subvectors $\vec{x}^{(m)}$, and $Y$ be the distribution of distances within that table. I.e.:
\begin{align}
    p(Y = y) \triangleq \int_{Q, X} p(\vec{q}^{(m)}, \vec{x}^{(m)})I\{\boldsymbol{\delta} \left( \vec{q}^{(m)}, \vec{x}^{(m)} \right) = y\}
\end{align}
We seek to learn a table-specific quantization function $\beta_m: \mathbb{R} \rightarrow \{0,\ldots,255\} $ that minimizes the quantization error. For computational efficiency, we constrain $\beta_m(y)$ to be of the form:
\begin{align}
    \beta_m(y) = \max(0, \min(255, \floor*{ay - b}))
\end{align}
for some constants $a$ and $b$. Formally, we seek values for $a$ and $b$ that minimize:
\begin{align}
    E_Y[(\hat{y} - y)^2]
\end{align}
where $\hat{y} \triangleq (\beta_m(y) + b)/a$ is termed the \textit{reconstruction} of $y$.
$Y$ can be an arbitrary distribution (though we assume it has finite mean and variance) and the value of $\beta_m(y)$ is constrained to a finite set of integers, so there is not an obvious solution to this problem.

We propose to set $b = F^{-1}(\alpha)$, $a = 255 / (F^{-1}(1 - \alpha) - b)$ for some suitable $\alpha$, where $F^{-1}$ is the inverse CDF of $Y$, estimated empirically. That is, we set $a$ and $b$ such that the $\alpha$ and $1 - \alpha$ quantiles of $Y$ are mapped to 0 and 255. Because both $F^{-1}(\alpha)$ and the loss function are cheap to compute, we can find a good $\alpha$ at training time with a simple grid search. In our experiments, we search over the values $\{0, .001, .002, .005, .01, .02, .05, .1\}$. In practice, the chosen $\alpha$ tends to be among the smaller values, consistent with the observation that loss from extreme values of $y$ is more costly than reduced granularity in representing typical values of $y$.

To quantize multiple lookup tables, we learn a $b$ value for each table and set $a$ based on the CDF of the aggregated distances $Y$ across all tables. We cannot learn table-specific $a$ values because this would amount to weighting distances from each table differently. The $b$ values can be table-specific because they sum to one overall bias, which is known at the end of training time and can be corrected for.


In summary, Bolt is an extension of product quantization with 1) fast encoding speed stemming from small codebooks, and 2) fast distance computations stemming from adaptively quantized lookup tables and efficient use of hardware.

\subsection{Theoretical Guarantees}

Due to space constraints, we state the following without proof. Supporting materials, including proofs and additional bounds, can be found on Bolt's website (see Section~\ref{sec:results}). 
Throughout the following, let $b_{min} \triangleq F^{-1}(\alpha)$, $b_{max} \triangleq F^{-1}(1 - \alpha)$, $\Delta \triangleq \frac{b_{max} - b_{min} }{ 256 }$, and $\sigma_Y \triangleq \sqrt{\Var[Y]}$. Furthermore, let the tails of $Y$ be drawn from any Laplace, Exponential, Gaussian, or subgaussian distribution, where the tails are defined to include the intervals $(-\inf, b_{min}]$ and $[b_{max}, \inf)$.

\begin{lemma}
$b_{min} \le y \le b_{max} \implies |y - \hat{y}| < \Delta$.
\end{lemma}

\begin{lemma}
For all $\eps > \Delta$, \hspace{1mm} $p(|y - \hat{y}| > \eps) <$
\begin{align}
        \frac{1}{\sigma_Y} \left(
            e^{-(b_{max}- E[Y]) / \sigma_Y}
            + e^{-(E[Y] - b_{min}) / \sigma_Y}
        \right)e^{-\eps / \sigma_Y}
\end{align}
\end{lemma}

We now bound the overall errors in dot products and Euclidean distances. First, regardless of the distributions of $\vec{q}$ and $\vec{x}$, the following hold:

\begin{lemma}
$\abs{\q^\top \x - \q^\top \xhat} < \norm{\q} \cdot \norm{\x - \xhat}$
\end{lemma}

\begin{lemma}
$\abs{\norm{\q - \x} - \norm{\q - \xhat} } < \norm{\x - \xhat}$
\end{lemma}

\noindent Using these bounds, it is possible to obtain tighter, probabilistic bounds using Hoeffding's inequality.

\begin{definition}[Reconstruction]
Let $\mat{C}$ be the set of codebooks used to encode $\vec{x}$. Then the vector obtained by replacing each $\vec{x}^{(m)}$ with its nearest centroid in codebook $\mathcal{C}_m$ is the \textit{reconstruction} of $\vec{x}$, denoted $\hat{\vec{x}}$.
\end{definition}

\begin{lemma}
Let $\r^{(m)} \triangleq \x^{(m)} - \xhat^{(m)}$, and assume that the values of $\norm{ \r^{(m)} }$ are independent for all $m$. Then: 
\begin{align} \label{eq:dot_indep}
    p(\abs{\q^\top \x - \q^\top \xhat} \ge \eps) \le 2\exp \left( \frac{-\eps^2}{
        2 \sum_{m=1}^M (\norm{\q^{(m)}} \cdot \norm{ \r^{(m)} }))^2
    }\right)
\end{align}
\end{lemma}


\begin{lemma}
Let $\r^{(m)} \triangleq \x^{(m)} - \xhat^{(m)}$, and assume that the values of $\norm{\q^{(m)} - \x^{(m)} }^2 - \norm{\q^{(m)} - \xhat^{(m)} }^2$ are independent for all $m$. Then: 
\begin{align} \label{eq:l2_indep}
    p(\abs{ \norm{\q - \x}^2 - \norm{\q - \xhat}^2 } > \eps) \le
        2\exp \left( \frac{-\eps^2}{
            2 \sum_{m=1}^M \norm{\r^{(m)}}^4
        }\right)
\end{align}
\end{lemma}

\section{Experimental Results} \label{sec:results}

To assess Bolt's effectiveness, we implemented both it and comparison algorithms in C++ and Python. All of our code and raw results are publicly available on the Bolt website.\footnote{https://github.com/dblalock/bolt} This website also contains experiments on additional datasets, as well as thorough documentation of both our code and experimental setups. All experiments use a single thread on a 2013 Macbook Pro with a 2.6GHz Intel Core i7-4960HQ processor. 

The goals of our experiments are to show that 1) Bolt is extremely fast at encoding vectors and computing scalar reductions, both compared to similar algorithms and in absolute terms; and 2) Bolt achieves this speed at little cost in accuracy compared to similar algorithms. To do the former, we record its throughput in encoding and computing reductions. To do the latter, we measure its accuracy in retrieving nearest neighbors, as well as the correlations between the reduced values it returns and the true values. Because they are by far the most benchmarked scalar reductions in related work and are widely used in practice, we test Bolt only on the Euclidean distance and dot product. Because of space constraints, we do not compare Bolt's distance table quantization method to possible alternatives, instead simply demonstrating that it yields no discernible loss of accuracy compared to exact distance tables.

For all experiments, we assess Bolt and the comparison methods using the commonly-employed encoding sizes of 8B, 16B, and 32B to characterize the relationships between space, speed, and accuracy. 

All reported timings and throughputs are the best of 5 runs, averaged over 10 trials (i.e., the code is executed 50 times). We use the best in each trial, rather than average, since this is standard practice in performance benchmarking. Because there are no conditional branches in either Bolt or the comparison algorithms (when implemented efficiently), all running times depend only on the sizes of the database and queries, not their distributions; consequently, we report timing results on random data.

\vspace{-2mm}
\subsection{Datasets}

For assessing accuracy, we use several datasets widely used to benchmark Multi-Codebook Quantization (MCQ) algorithms:
\begin{itemize}[leftmargin=4mm]
\item \textbf{Sift1M} \cite{pq} --- 1 million 128-dimensional SIFT \cite{sift} descriptors of images. Sift1M vectors tend to have high correlations among many dimensions, and are therefore highly compressible. This dataset has a predefined query/train database/test database split, consisting of 10,000 query vectors, 100,000 training vectors, and 1 million database vectors.
\item \textbf{Convnet1M} \cite{stackedQuantizers} --- 1 million 128-dimensional Convnet descriptors of images. These vectors have some amount of correlation, but less than Sift1M. It has a query/train/test split matching that of Sift1M.
\item \textbf{LabelMe22k} \cite{minimalLossHashing} --- 22,000 512-dimensional GIST descriptors of images. Like Sift, it has a great deal of correlation between many dimensions. It only has a train/test split, so we follow \cite{lsq, cq} and use the 2,000-vector test set as the queries and the 20,000 vector training set as both the training and test database.
\item \textbf{MNIST} \cite{mnist} --- 60,000 28x28-pixel greyscale images, flattened to 784-dimensional vectors. This dataset is sparse and has high correlations between various dimensions. Again following \cite{lsq} and \cite{cq}, we split it the same way as the LabelMe dataset.
\end{itemize}

For all datasets, we use a portion of the training database as queries when learning Bolt's lookup table quantization.


\vspace{-2mm}
\subsection{Comparison Algorithms}

Our comparison algorithms include MCQ methods that have high encoding speeds ($\ll 1$ms / vector on a CPU). If encoding speed is not a design consideration or is dominated by a need for maximal compression, methods such as GRVQ \cite{grvq} or LSQ \cite{lsq} are more appropriate than Bolt.\footnote{Although Bolt \textit{might} still be desirable for its high query speed even if encoding speed is not a consideration.}

Our primary baselines are Product Quantization (PQ) \cite{pq} and Optimized Product Quantization (OPQ) \cite{opq}, since they offer the fastest encoding times. There are several algorithms that extend these basic approaches by adding indexing methods \cite{lopq, NOIMI}, or more sophisticated training-time optimizations \cite{googleMips, pairQ, polysemous}, but since these extensions are compatible with our own work, we do not compare to them. We compare only to versions of PQ and OPQ that use 8 bits per codebook, since this is the setting used in all related work of which we are aware; we do not compare to using 4 bits, as in Bolt, since this both reduces their accuracy and increases their computation time. Note that, because the number of bits per codebook is fixed in all methods, varying the encoded representation size means varying the number of codebooks.

We do not to compare to binary embedding methods in terms of accuracy as they are known to yield much lower accuracy for a given code length than MCQ methods \cite{hashingSimilaritySurvey, opq} and, as we show, are also slower in computing distances than Bolt. 

We have done our best to optimize the implementations of the comparison algorithms, and find that we obtain running times superior to those described in previous works. For example, \cite{stackedQuantizers} reports encoding roughly 190,000 128-dimensional vectors per second with PQ, while our implementation encodes nearly 300,000.

As a final comparison, we include a modified version of Bolt, \textit{Bolt No Quantize}, in our accuracy experiments. This version does not quantize the distance lookup tables. It is not a useful algorithm since it sacrifices Bolt's high speed, but it allows us to assess whether our codebook quantization reduces accuracy.

\subsection{Encoding Speed}

\begin{figure}[h]
\begin{center}
\includegraphics[width=\linewidth]{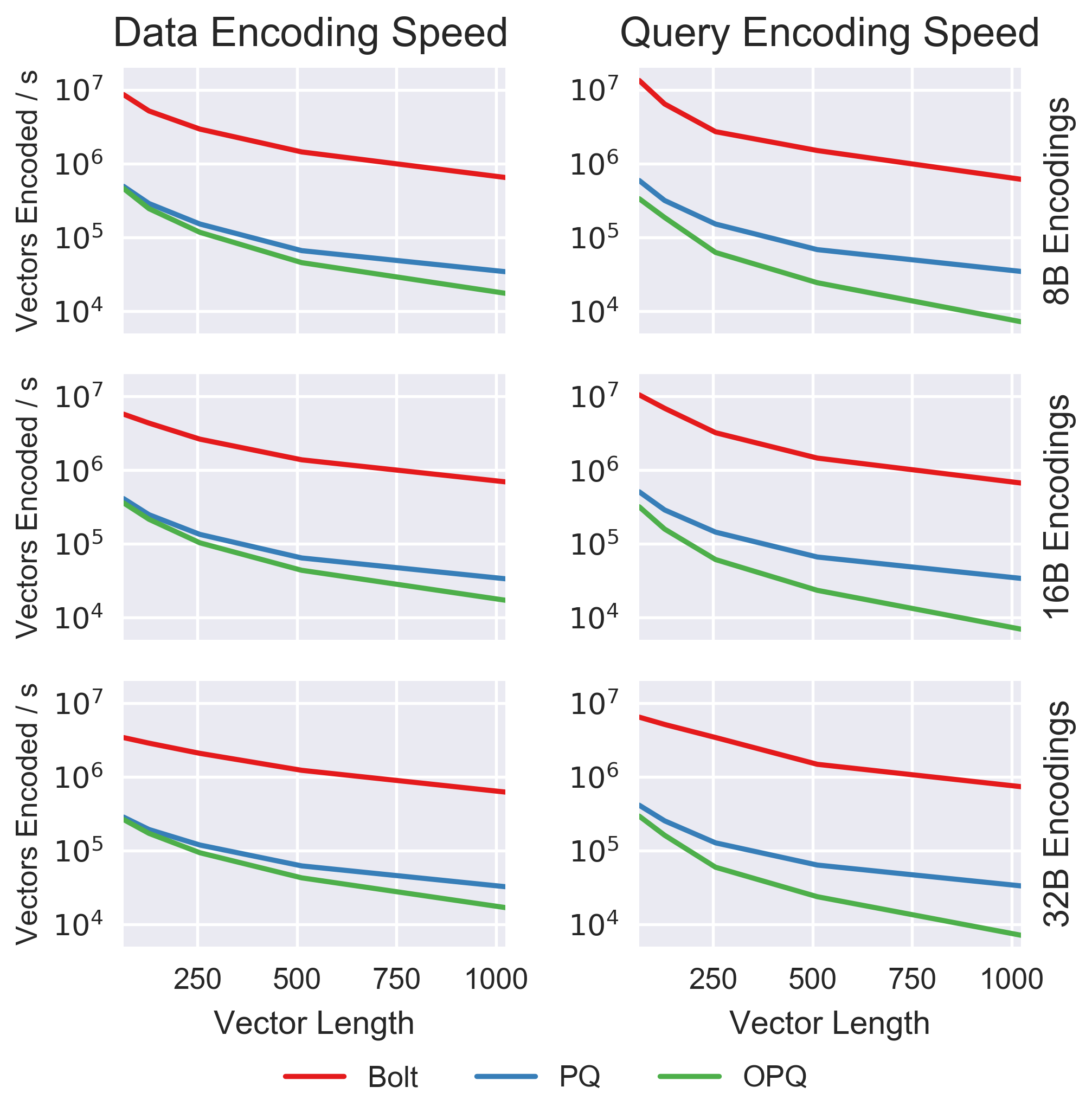}
\vspace*{-2mm}
\caption{Bolt encodes both data and query vectors significantly faster than similar algorithms.}
\label{fig:encoding_speeds}
\end{center}
\end{figure}

Before a vector quantization method can compute approximate distances, it must first encode the data. We measured how many vectors each algorithm can encode per second as a function of the vectors' length. As shown in Figure~\ref{fig:encoding_speeds}.\textit{left}, Bolt can encode data vectors over $10\times$ faster than PQ, the fastest comparison. Encoding $5$ million 128-dimensional vectors of $4B$ floats per second (top left plot) translates to an encoding speed of $2.5$GB/s. For perspective, this encoding rate is sufficient to encode the entire Sift1M dataset of 1 million vectors in 200ms, and the Sift1B dataset of 1 billion vectors in 200s. This rate is also much higher than that of high-speed (but general-purpose) compression algorithms such as Snappy \cite{snappy}, which reports an encoding speed of 250MB/s.

Similarly, Bolt can compute the distance matrix constituting a query's encoding at over 6 million queries/s (top right plot), while PQ obtains less than $350,000$ queries/s. Both of these numbers are sufficiently high that encoding the query is unlikely to be a bottleneck in computing distances to it.

\vspace{-2mm}
\subsection{Query Speed}

Much of the appeal of MCQ methods is that they allow fast computation of approximate distances and similarities directly on compressed data. We assessed various algorithms' speeds in computing Euclidean distances from a set of queries to each vector in a compressed dataset. We do not present results for other distances and similarities since they only affect the computation of queries' distance matrices and therefore have speeds nearly identical to those shown here. In all experiments, the number of compressed data vectors $N$ is fixed at $100,000$ and their dimensionality is fixed at $256$.

We compare Bolt not only to other MCQ methods, but also to other methods of computing distances that might serve as reasonable alternatives to using MCQ at all. These methods include:
\begin{itemize}[leftmargin=4mm]
    \item \textit{Binary Embedding}. As mentioned in Section~\ref{sec:relatedWork}, the current fastest method of obtaining approximate distances over compressed vectors is to embed them into Hamming space and use the \texttt{popcount} instruction to quickly compute Hamming distances between them.
    \item \textit{Matrix Multiplies}. Given the norms of query and database vectors, Euclidean distances can be computed using matrix-vector multiplies. When queries arrive quickly relative to the latency with which they must be answered, multiple queries can be batched into a matrix. Performing one matrix multiply is many times faster than performing individual matrix-vector multiplies. We compare to batch sizes of $1$, $256$, and $1024$.
\end{itemize}

\begin{figure}[h]
\begin{center}
\includegraphics[width=\linewidth]{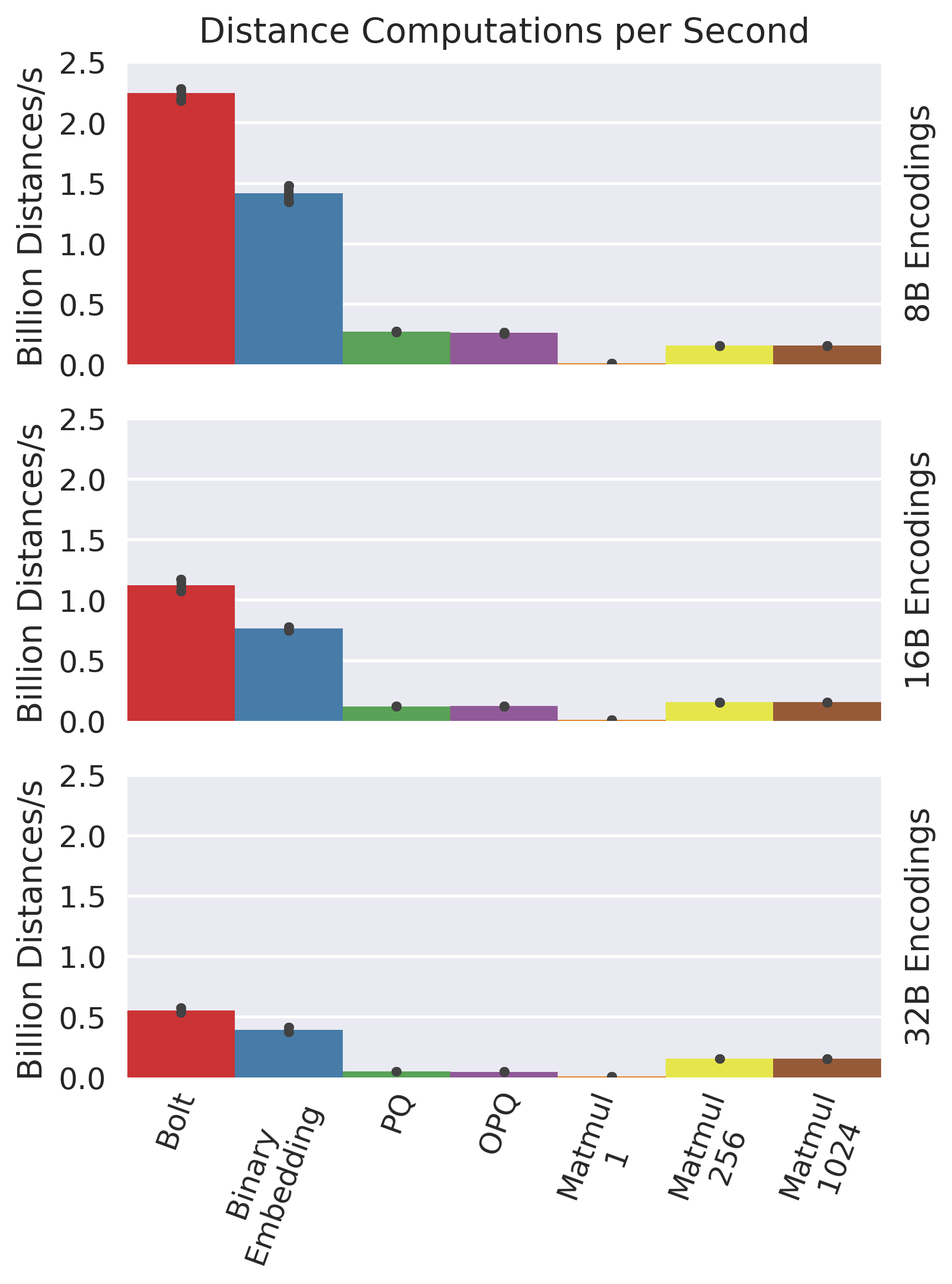}
\vspace*{-3mm}
\caption{Bolt can compute the distances/similarities between a query and the vectors of a compressed database up to $10\times$ faster than other MCQ algorithms. It is also faster than binary embedding methods, which use the hardware \texttt{popcount} instruction, and matrix-vector multiplies using batches of 1, 256, or 1024 vectors.}
\label{fig:query_speeds}
\end{center}
\end{figure}

Bolt computes Euclidean distances up to ten times faster than any other MCQ algorithm and significantly faster than binary embedding methods can compute Hamming distances (Figure~\ref{fig:query_speeds}). Its speedup over matrix multiplies depends on the batch size and number of bytes used in MCQ encoding. When it is not possible to batch multiple queries (\textit{Matmul 1}), Bolt 8B is roughly $250\times$ faster, Bolt 16B is $140\times$ faster, and Bolt 32B is $60\times$ faster (see website for exact timings). When hundreds of queries can be batched (\textit{Matmul 256}, \textit{Matmul 1024}), these numbers are reduced to roughly $13\times$, $7\times$, and $3\times$.

\vspace{-2mm}

\begin{figure}[h]
\begin{center}
\includegraphics[width=\linewidth]{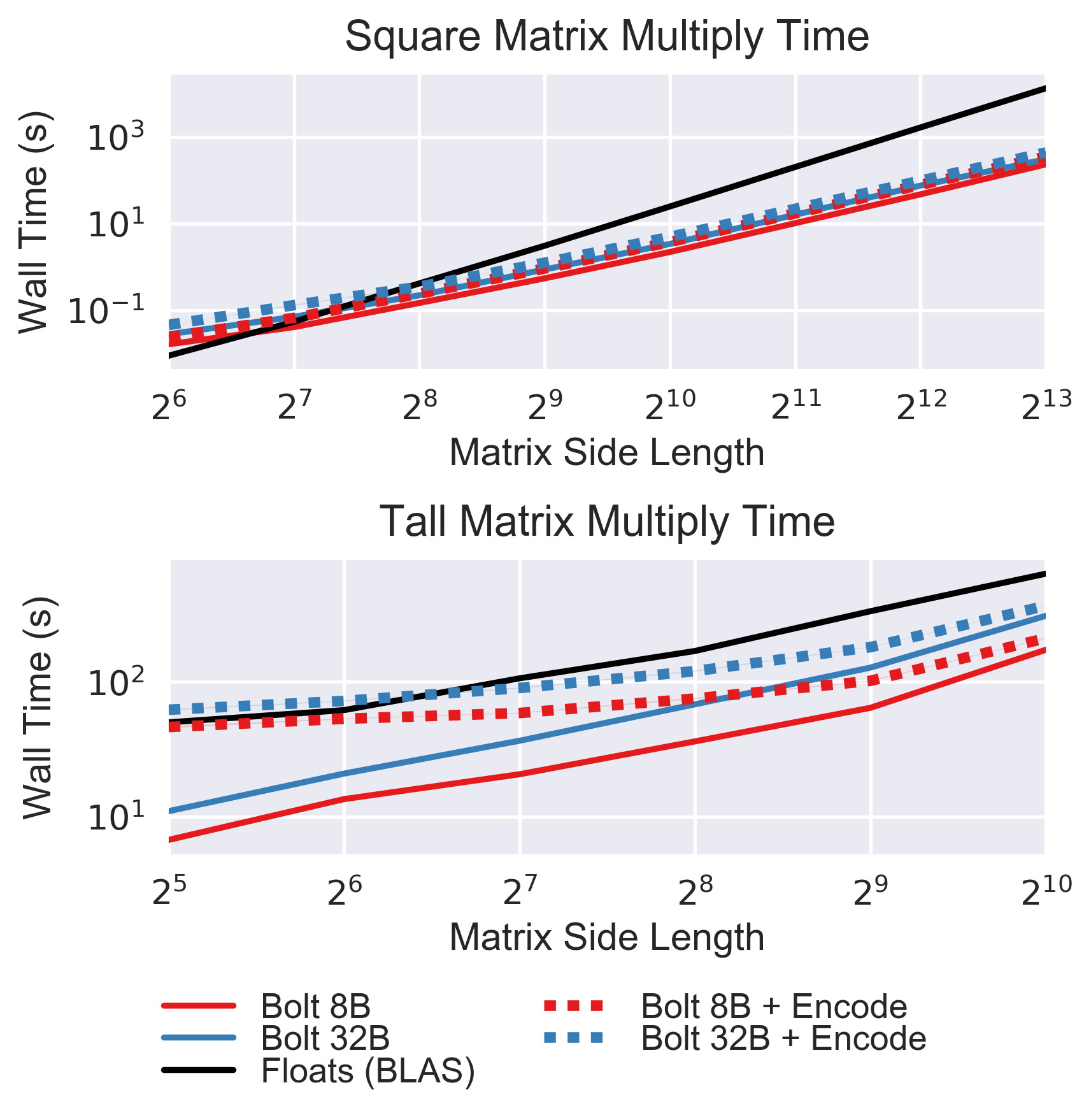}
\vspace*{-3.5mm}
\caption{Using a naive nested loop implementation, Bolt can compute (approximate) matrix products faster than optimized matrix multiply routines. Except for small matrices, Bolt is faster even when it must encode the matrices from scratch as a first step.}
\label{fig:matmul_speed}
\end{center}
\end{figure}

Because matrix multiplies are so ubiquitous in data mining, machine learning, and many other fields, we compare Bolt to matrix multiplication in more detail. In Figure~\ref{fig:matmul_speed}, we profile the time that Bolt and a state-of-the-art BLAS implementation \cite{eigen} take to do matrix multiplies of various sizes. Bolt computes matrix multiplies by treating each row of the first matrix as a query, treating the second matrix as the database, and iteratively computing the inner products between each query and all database vectors. This nested-loop implementation is not optimal, but Bolt is still able to outperform BLAS.

In Figure~\ref{fig:matmul_speed}\textit{.top}, we multiply two square matrices of varying sizes, which is the optimal scenario for most matrix multiply algorithms. For small matrices, the cost of encoding one matrix as the database is too high for Bolt to be faster. For larger matrices, this cost is amortized over many more queries, and Bolt becomes faster. When the database matrix is already encoded, Bolt is faster for almost all matrix sizes, even using 32B encodings. Note, though, that this comparison ceases to be fair for especially large matrices (e.g. $4096 \times 4096$) since encoding so many dimensions accurately would almost certainly require more than 32B.

In Figure~\ref{fig:matmul_speed}\textit{.bottom}, we multiply a 100,000 $\times$ 256 matrix by a $256 \times n$ matrix. Bolt uses the rows of the former matrix as the database and the columns of the latter as the queries. Again, Bolt is slower for small matrices when it must first encode the database, but always faster for larger ones or when it does not need to encode the database. Because only the number of queries is changing and not the dimensionality of each vector, longer encodings would not be necessary for the larger matrices.

\vspace{-2mm}

\vspace*{-4mm}
\subsection{Nearest Neighbor Accuracy}

\begin{figure}[h]
\begin{center}
\includegraphics[width=\linewidth]{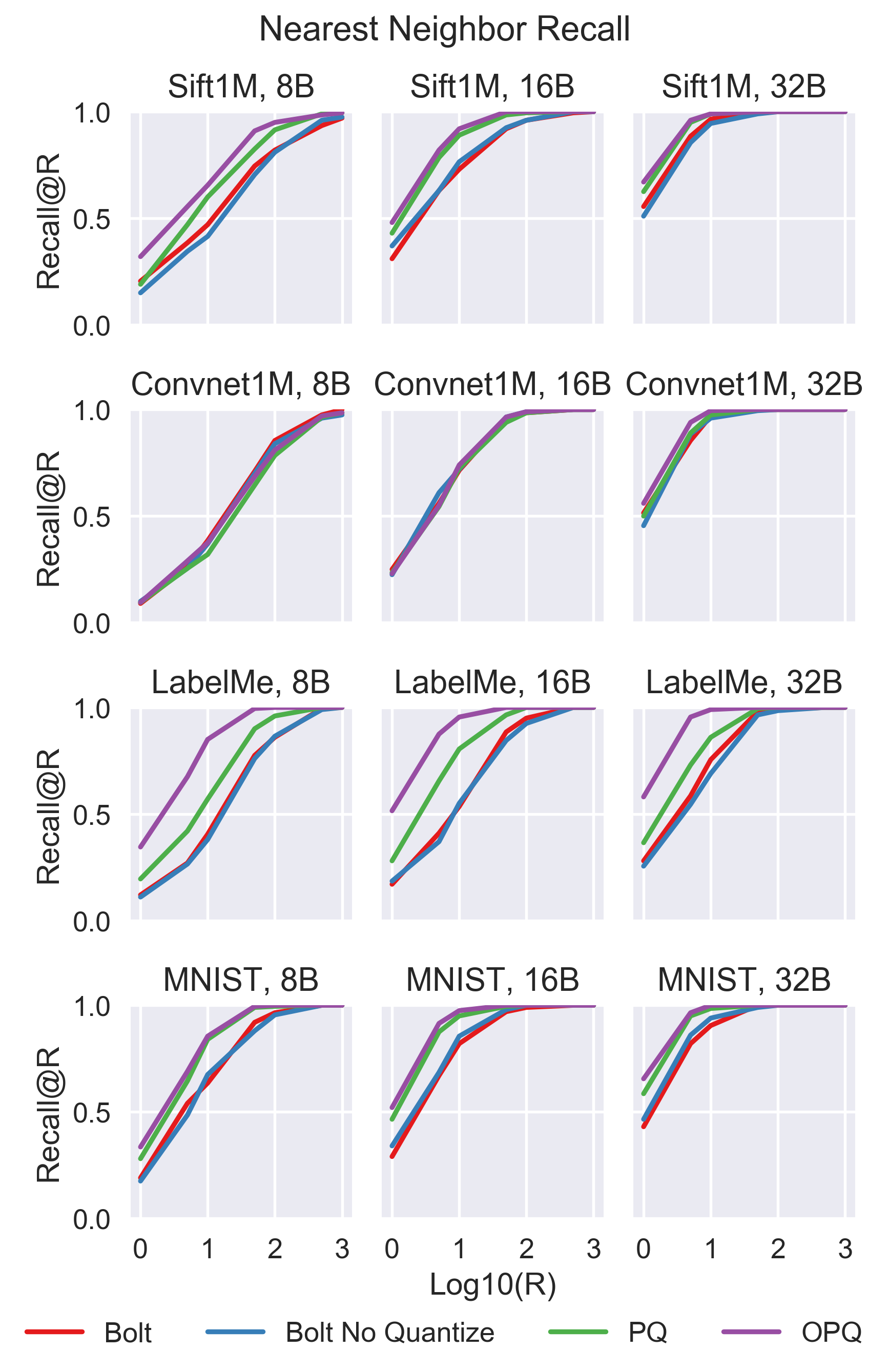}
\vspace*{-3mm}
\caption{Compared to other MCQ algorithms, Bolt is slightly less accurate in retrieving the nearest neighbor for a given encoding length.}
\label{fig:nn_acc}
\end{center}
\end{figure}

The most common assessment of MCQ algorithms' accuracy is their Recall@R. This is defined as the fraction of the queries $\vec{q}$ for which the true nearest neighbor in Euclidean space is among the top $R$ points with smallest approximate distances to $\vec{q}$. This is a proxy for how many points would likely have to be reranked in a retrieval context when using an approximate distance measure to generate a set of candidates. As shown in Figure~\ref{fig:nn_acc}, Bolt yields slightly lower accuracy for a given encoding length than other (much slower) MCQ methods. The nearly identical curves for Bolt and Bolt No Quantize suggest that our proposed lookup table quantization introduces little or no error. 

The differences across datasets can be explained by their varying dimensionalities and the extent to which correlated dimensions tend to be in the same subspaces. In the Sift1M dataset, adjacent dimensions are highly correlated, but they are also correlated with other dimensions slightly farther away. This first characteristic allows all algorithms to perform well, but the second allows PQ and OPQ to perform even better thanks to their smaller numbers of larger codebooks. Having fewer codebooks means that the subspaces associated with each are larger (i.e., more dimensions are quantized together), allowing mutual information between them to be exploited. Bolt, with its larger number of smaller codebooks, must quantize more sets of dimensions independently, which does not allow it to exploit this mutual inforation. Much the same phenomena explain the results on MNIST.

For the LabelMe dataset, the correlations between dimensions tend to be even more diffuse, with small correlations spanning dimensions belonging to many subspaces. This is less problematic for OPQ, which learns a rotation such that correlated dimensions tend to be placed in the same subspaces. PQ and Bolt, which lack the ability to rotate the data, have no such option, and so are unable to encode the data as effectively.

Finally, for the Convnet1M dataset, most of the correlated dimensions tend to be immediately adjacent to one another, allowing all methods to perform roughly equally.

\vspace{-4mm}
\subsection{Accuracy in Preserving Distances and Dot Products}

\vspace{1mm}

\begin{figure}[h]
\begin{center}
\includegraphics[width=\linewidth]{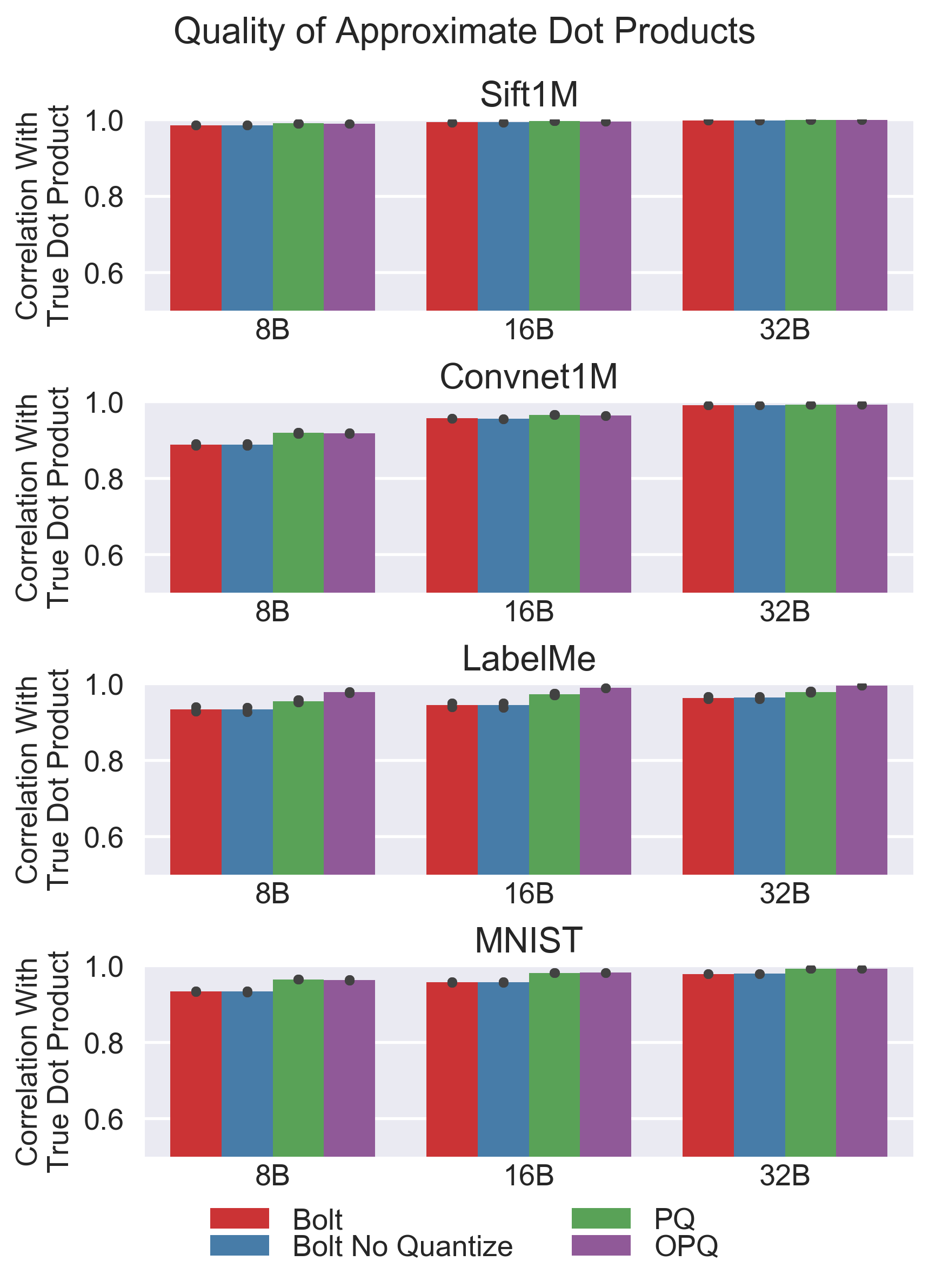}
\vspace*{-2mm}
\caption{Bolt dot products are highly correlated with true dot products, though slightly less so than those from other MCQ algorithms.}
\label{fig:dotprod_distortion}
\end{center}
\end{figure}

The Recall@R experiment characterizes how well each algorithm preserves distances to highly similar points, but not whether distances in general tend to be preserved. To assess this, we computed the correlations between the true dot products and approximate dot products for Bolt and the comparison algorithms. Results for Euclidean distances are similar, so we omit them. As Figure~\ref{fig:dotprod_distortion} illustrates, Bolt is again slightly less accurate than other MCQ methods. In absolute terms, however, it consistently exhibits correlations with the true dot products above $.9$, and often near $1.0$. This suggests that its approximations could reliably be used instead of exact computations when slight errors are permissible.

For example, if one could tolerate a correlation of $.9$, one could use Bolt 8B instead of dense vectors of 4B floats and achieve dramatic speedups, as well as compression ratios of $64\times$ for SIFT1M and Convnet1M, $256\times$ for LabelMe, and $392\times$ for MNIST. If one required correlations of $.95$ or more, one could use Bolt 32B and achieve slightly smaller speedups with compression ratios of $16\times$, $64\times$, and $98\times$.

\vspace{-1mm}
\section{Summary} \label{sec:conclusion}

We describe Bolt, a vector quantization algorithm that rapidly compresses large collections of vectors and enables fast computation of approximate Euclidean distances and dot products directly on the compressed representations. Bolt both compresses data and computes distances and dot products up to $10\times$ faster than existing algorithms, making it advantageous both in read-heavy and write-heavy scenarios. Its approximate computations can be over $100\times$ faster than the exact computations on the original floating-point numbers, while maintaining correlations with the true values of over $.95$. Moreover, at this level of correlation, Bolt can achieve $10$-$200\times$ compression or more. These attributes make Bolt ideal as a subroutine in algorithms that are amenable to approximate computations, such as nearest neighbor or maximum inner product searches.

It is our hope that Bolt will be used in many production systems to greatly reduce storage and computation costs for large, real-valued datasets.

\vspace{-1mm}


\bibliographystyle{ACM-Reference-Format}
\bibliography{doc}

\end{document}